\documentclass[]{AO4ELT}  

\usepackage{microtype}
\usepackage{biblatex}
\usepackage{amsmath,amsfonts,amssymb}
\usepackage{graphicx}
\usepackage{pst-all} 
\usepackage[colorlinks=true, allcolors=blue]{hyperref}
\usepackage{float}
\makeatletter         
\def\@maketitle{
\includegraphics[width = 170mm]{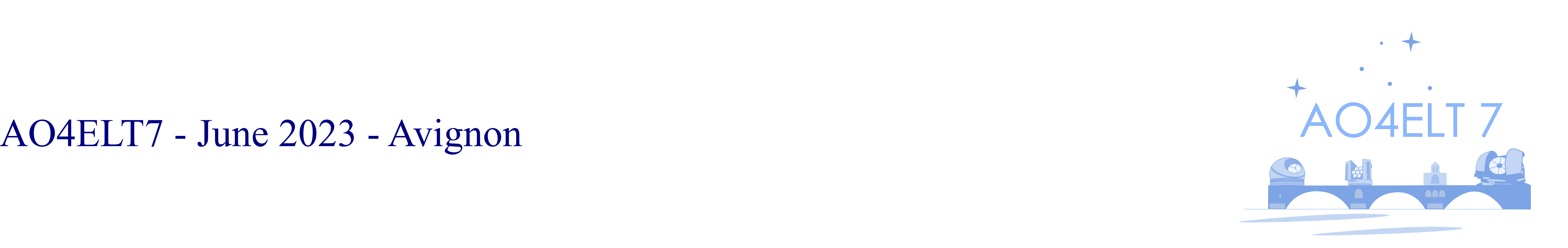}\\[8ex]
\begin{center}
{\Huge \bfseries \sffamily \@title }\\[4ex] 
{\Large  \@author}\\[4ex] 
\@date
\end{center}}

\title{GMagAO-X: A First Light Coronagraphic Adaptive Optics System for the GMT}

\author[a]{Maggie Kautz}
\author[b]{Jared Males}
\author[b]{Laird Close}
\author[b]{Sebastiaan Haffert}
\author[a,b]{Olivier Guyon}
\author[d]{Alexander Hedglen}
\author[b]{Victor Gasho}
\author[b]{Olivier Durney}
\author[b]{Jamison Noenickx}
\author[b]{Adam Fletcher}
\author[b]{Fernando Coronado}
\author[b]{John Ford}
\author[b]{Tom Connors}
\author[b]{Mark Sullivan}
\author[b]{Tommy Salanski}
\author[b]{Doug Kelly}
\author[c]{Richard Demers}
\author[c]{Antonin Bouchez}
\author[c]{Breann Sitarski}
\author[c]{Patricio Schurter}
\affil[a]{James C. Wyant College of Optical Sciences, University of Arizona, 1630 E University Blvd, Tucson, AZ, USA 85721}
\affil[b]{Steward Observatory, University of Arizona, 933 N Cherry Ave, Tucson, AZ, USA 85719}
\affil[c]{Giant Magellan Telescope Organization, 300 N Lake Ave 14th Floor, Pasadena, CA, USA 91101}
\affil[d]{Northrop Grumman Corporation, 600 S Hicks Rd, Rolling Meadows, IL, USA 60008}

\authorinfo{Further author information: (Send correspondence to M.K.)\\M.K.: E-mail: maggiekautz@arizona.edu}

\begin{document} 
\maketitle
\begin{abstract}
GMagAO-X is a visible to NIR extreme adaptive optics (ExAO) system that will be used at first light for the Giant Magellan Telescope (GMT). GMagAO-X is designed to deliver diffraction-limited performance at visible and NIR wavelengths (6 to 10 mas) and contrasts on the order of 10\textsuperscript{-7}. The primary science case of GMagAO-X will be the characterization of mature, and potentially habitable, exoplanets in reflected light. GMagAO-X employs a woofer-tweeter system and includes segment phasing control. The tweeter is a 21,000 actuator segmented deformable mirror (DM), composed of seven individual 3,000 actuator DMs. This new ExAO framework of seven DMs working in parallel to produce a 21,000 actuator DM significantly surpasses any current or near future actuator count for a monolithic DM architecture. Bootstrapping, phasing, and high order sensing are enabled by a multi-stage wavefront sensing system. GMT’s unprecedented 25.4 m aperture composed of seven segments brings a new challenge of co-phasing massive mirrors to 1/100\textsuperscript{th} of a wavelength. The primary mirror segments of the GMT are separated by large $>$30 cm gaps so there will be fluctuations in optical path length (piston) across the pupil due to vibration of the segments, atmospheric conditions, etc. We have developed the High Contrast Adaptive-optics Testbed (HCAT) to test new wavefront sensing and control approaches for GMT and GMagAO-X, such as the holographic dispersed fringe sensor (HDFS), and the new ExAO parallel DM concept for correcting aberrations across a segmented pupil. The CoDR for GMagAO-X was held in September 2021 and a preliminary design review is planned for early 2024. In this paper we will discuss the science cases and requirements for the overall architecture of GMagAO-X, as well as the current efforts to prototype the novel hardware components and new wavefront sensing and control concepts for GMagAO-X on HCAT.  
\end{abstract}


\section{INTRODUCTION}
\label{sec:intro}  

The Giant Magellan Telescope will be one of the first in a new line of extremely large telescopes (ELTs). The Gregorian telescope will be comprised of seven 8.4 m segments, with a seven segment adaptive secondary. The full diameter is 25.4 meters with a resolving power of a 24.5 meter telescope. This unprecedented size will allow for extremely high resolution optical to near-infrared images.

This new era of extremely large telescopes (ELTs) will enable a huge leap forward in the characterization of extrasolar planets. The unprecedented large diameters (D) will scale the sensitivity as D\textsuperscript{4} and the angular resolution proportional to 1/D. The combination of these new capabilities with an extreme adaptive optics systems, like GMagAO-X, delivering diffraction-limited resolution will allow for direct imaging and spectroscopic characterization of exoplanets at closer separations than ever previously done.

    \begin{figure} [H]
\begin{center}
\begin{tabular}{c} 
\includegraphics[height=7cm]{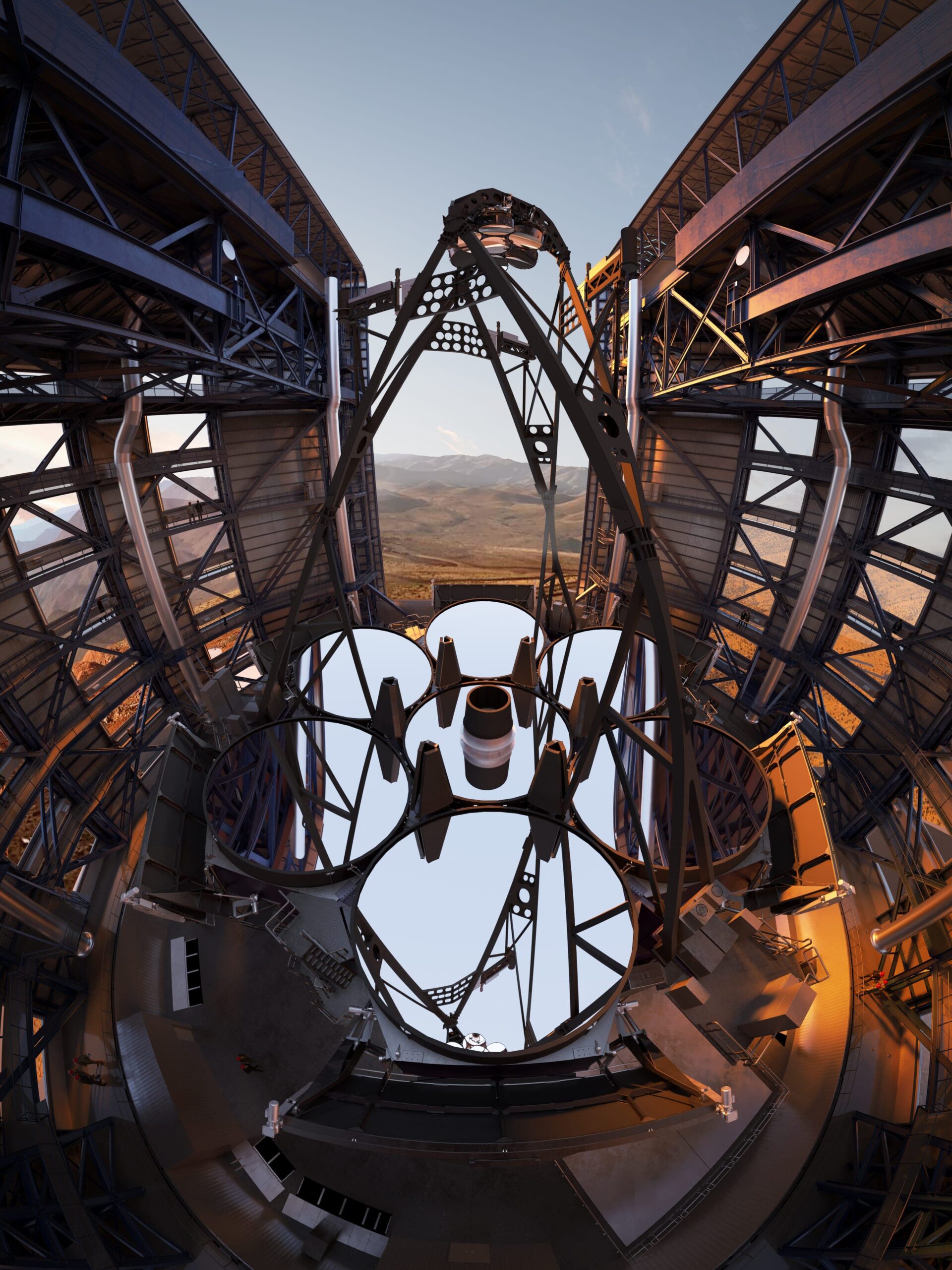}
\end{tabular}
\end{center}
\caption{Courtesy: Giant Magellan Telescope Organization}
\label{fig:GMT} 
    \end{figure}

The Magellan Extreme Adaptive Optics system (MagAO-X), is an active extreme AO system that operates at the 6.5-meter Magellan Clay Telescope at the Las Campanas Observatory in Chile\cite{males_magao-x_2020, close_optical_2018}. It employs a woofer-tweeter system with a 2,040-actuator deformable mirror (DM) operating up to 3.63kHz commanded by a pyramid wavefront sensor. This has been very successful, producing high-resolution, high-Strehl images\cite{males_magao-x_2022}. GMagAO-X, the Giant Magellan Extreme Adaptive Optics system, is a next generation instrument designed on similar principles but for the much larger and segmented Giant Magellan Telescope\cite{close_concept_2020, males_conceptual_2022}. GMagAO-X held its Conceptual Design Review (CoDR) in September 2021 and is moving towards a preliminary design review (PDR) in spring 2024, in hopes to be included as a first light instrument at the GMT.

\section{Science Goals}
The diffraction limited resolution of GMT will be 4.1 mas at V band and 13.2 mas at H band. This extraordinary angular resolution will allow for an expansion upon today's science and unlock a variety of new science cases. One of the main objectives will be to characterize of mature, temperate, and potentially habitable, exoplanets using reflected light. Reflected light shorter than H band will allow for a push to older, smaller, and cooler planets. Using the favorable angular resolution of bluer wavelength light, direct imaging of planets with smaller separations from their host star will be possible. The primary science goals of GMagAO-X are stated below \cite{males_conceptual_2022}.

\begin{enumerate}
  \item \textbf{Search For Life} [new science]
    \begin{itemize}
        \item Search for life on terrestrial exoplanets orbiting nearby stars
    \end{itemize}
  \item \textbf{Characterize Older Temperate Exoplanets} [new science] 
    \begin{itemize}
        \item Reflected light characterization of wide range of radii and mass
    \end{itemize}
  \item \textbf{Measure Orbit and Mass of Exoplanets} [today’s science, better]
    \begin{itemize}
        \item Feed to G-CLEF for Precision-RV 
        \item Identify new candidates 
        \item Obtain precise ephemerides for imaging targets
    \end{itemize}
  \item \textbf{Study Planet Formation at Low Mass and Small Separation} [today’s science, better] 
    \begin{itemize}
        \item Search for and characterize thermally self-luminous young planets (YJH) 
        \item Characterize forming planets through the H-alpha accretion signature 
    \end{itemize}
  \item \textbf{Circumstellar Disk Structure and Disk-Planet Interactions} [today’s science, better]
    \begin{itemize}
        \item Structure in scattered light at high spatial resolution
        \item Study disk-planet interactions
        \item Jets and outflows
    \end{itemize}
  \item \textbf{Refine Stellar Evolution Models} [today’s science, better]
    \begin{itemize}
        \item High spatial and spectral resolution characterization of benchmark binaries
    \end{itemize}
  \item \textbf{High Spatial Resolution Mapping} [today’s science, better]
    \begin{itemize}
        \item Moons, asteroids, stellar surfaces
    \end{itemize}
\end{enumerate}

\section{Instrument Overview}

GMagAO-X main components (adapted from Males et al. 2022)\cite{males_conceptual_2022}:
\begin{itemize}
  \item Global TTM: PI S-331
  \item Woofer DM: ALPAO DM 3228 (64 actuator diameter)
  \item VIS Pyramid WFS (R-I bands) 
  \item IR Pyramid WFS (Z-H bands) (switchable to Zernike WFS)
  \item Phase Control: 6X piezoelectric tip/tilt/piston PI S-325 stages
  \item Tweeter DM: 7X BMC 3000 actuator MEMS DMs
  \item Coarse Phasing Sensor: Holographic Dispersed Fringe Sensor (HDFS, using 10\% of J-H band light)
  \item Coronagraph NCP DM: BMC 3000 actuator MEMS DM
  \item Coronagraph: Lyot-architecture, supporting PIAACMC
  \item Science Instruments: Still being determined. Some V band to H band possibilities include:
    \begin{itemize}
      \item Dual-band imager for SDI
      \item Polarimeter in the visible and IR
      \item Low-resolution to high-resolution visible IFU (R=50 to R=15.000, similar to VIS-X)
      \item Low-resolution to medium-resolution IR IFU (R=100 to R=5000, similar to SCEXAO/CHARIS and GPI IFU)
      \item High-resolution visible spectroscopy by feeding G-CLEF with a fiber-feed
  \end{itemize}
\end{itemize}

GMagAO-X should be able to reach Strehls of $>$80$\%$ at 800 nm and raw contrasts on the order of 10\textsuperscript{-5}\cite{males_conceptual_2022}.

    \begin{figure} [H]
\begin{center}
\begin{tabular}{c} 
\includegraphics[height=5cm]{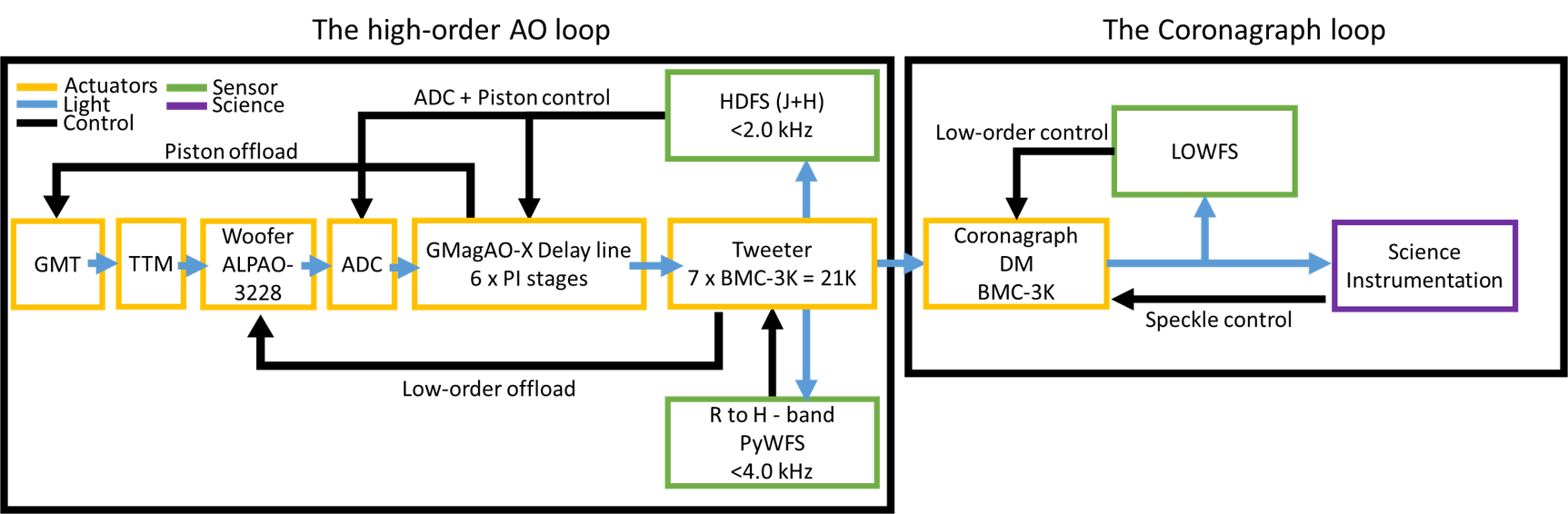}
\end{tabular}
\end{center}
\caption{Wavefront Sensing and Control Architecture}
\label{fig:wfs} 
    \end{figure}

\begin{table}[H]
\caption{Deformable Mirror Specifications Summary} 
\label{tab:hex_specs}
\begin{center}       
\resizebox{\textwidth}{!}{\begin{tabular}{|c|c|c|c|c|} 
\hline
\rule[-1ex]{0pt}{3.5ex}  \textbf{Mirror} & \textbf{Frequency} & \textbf{P2V Surface Stroke in low order modes} & \textbf{Pitch} & \textbf{Modes Corrected}  \\
\hline
\rule[-1ex]{0pt}{3.5ex}  Global FSM (PI S-331) & 1-2kHz (up to 10kHz) & 3-5mrad $\Theta$X $\Theta$Y & 25.4 m & 2  \\
\hline
\rule[-1ex]{0pt}{3.5ex}  Woofer (ALPAO 3k) & $\sim$1kHz & $\sim$7.5$\mu$m & 41 cm & 3,000   \\
\hline
\rule[-1ex]{0pt}{3.5ex}  Tweeter (7X BMC 3ks) & $\sim$2.63kHz (up to 12kHz) & $\sim$3.5$\mu$m & 14 cm & 21,000  \\
\hline
\rule[-1ex]{0pt}{3.5ex}  Tweeter Piston (6X PI S-325s) & 1kHz & $\pm$15$\mu$m Z, 5mrad  $\Theta$X $\Theta$Y & 8.4 m & 18   \\
\hline
\rule[-1ex]{0pt}{3.5ex}   NCPDM (BMC 3k) & $\sim$2.63kHz (up to 12kHz) & $\sim$3.5$\mu$m & 41 cm & 3,000  \\
\hline
\end{tabular}}
\end{center}
\end{table}

There will be a phase-induced amplitude apodization complex mask coronagraph (PIAACMC) utilized to suppress starlight. PIAACMCs are good for complex aperture telescopes and for achieving high inner working angles (IWA) with high throughput due to the lossless apodization\cite{guyon_high_2010}.

    \begin{figure}[H]
\centering
\begin{tabular}{c c}
\includegraphics[width=0.3\textwidth]{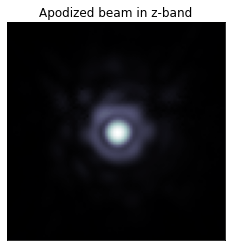} &
\includegraphics[width=0.3\textwidth]{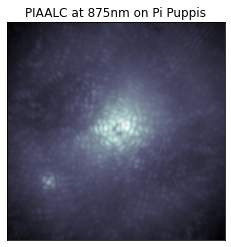} \\
\end{tabular}
\caption{(left) PIAA (right) PIAALC used on-sky with MagAO-X (images courtesy of Warren Foster's Masters Thesis in Optical Sciences at the University of Arizona\cite{foster_implementation_2023})}
\label{fig:PIAA}
    \end{figure}

The tweeter DM utilizing seven 3,000 actuator deformable mirrors is a novel wavefront sensing concept called the ``parallel DM''\cite{close_concept_2020, males_conceptual_2022, kautz_novel_2022}. The outer six segments of the GMT will each be assigned their own wavefront control line with one PI S-325 tip/tilt/piston stage for phasing and one 3,000 actuator MEMS DM for wavefront control. The inner segment will have its own 3k MEMS DM as well.

    \begin{figure}[H]
\centering
\begin{tabular}{c c}
\includegraphics[width=0.4\textwidth]{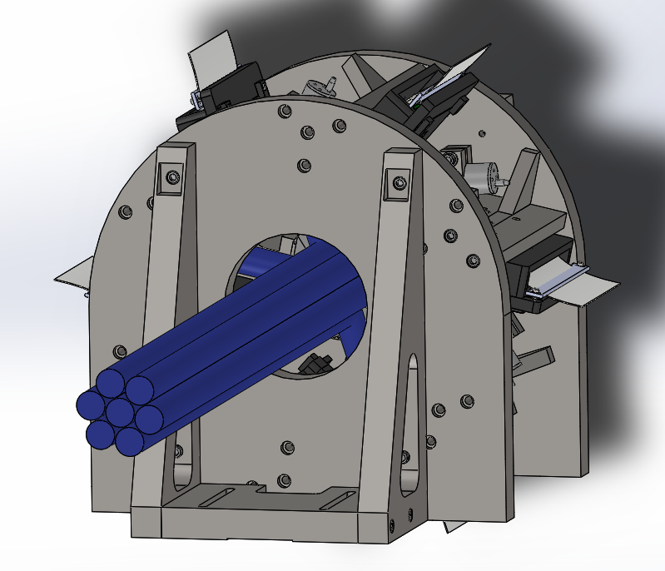} & \includegraphics[width=0.4\textwidth]{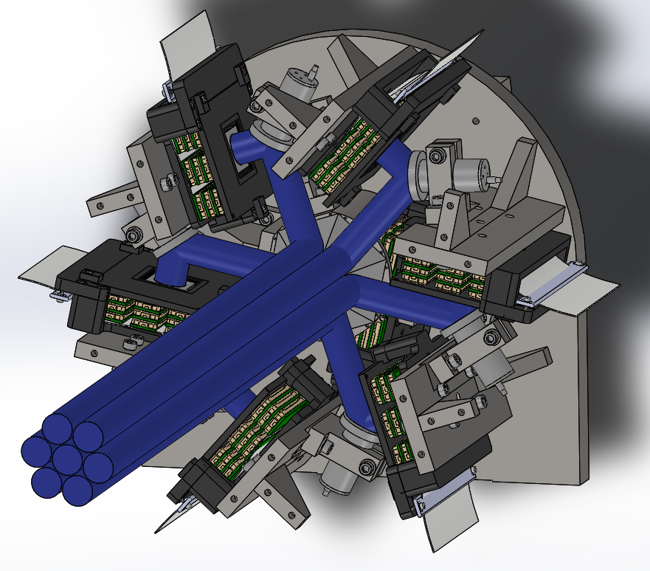} \\
\end{tabular}
\caption{CAD renderings of parallel DM}
\label{fig:Parallel_DM}
    \end{figure}

GMagAO-X will be mounted in the Gregorian Instrument Rotator (GIR) in a Folded Port (FP) and utilizes its own independent M3 mirror.

\clearpage

\section{Optomechanical Design Progress}

\subsection{Mechanical Design}

The heart of the instrument is an air-isolated two-tiered steel floating optical table. The tables are supported by the steel instrument support system (ISS).

    \begin{figure} [H]
\begin{center}
\begin{tabular}{c} 
\includegraphics[height=7cm]{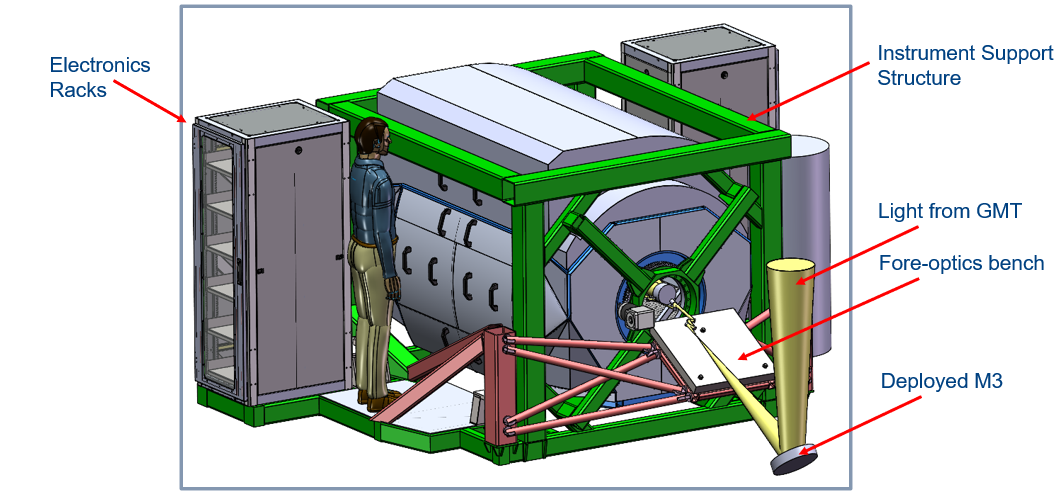}
\end{tabular}
\end{center}
\caption{CAD rendition of whole system}
\label{fig:system} 
    \end{figure}

    \begin{figure} [H]
\begin{center}
\begin{tabular}{c} 
\includegraphics[height=8cm]{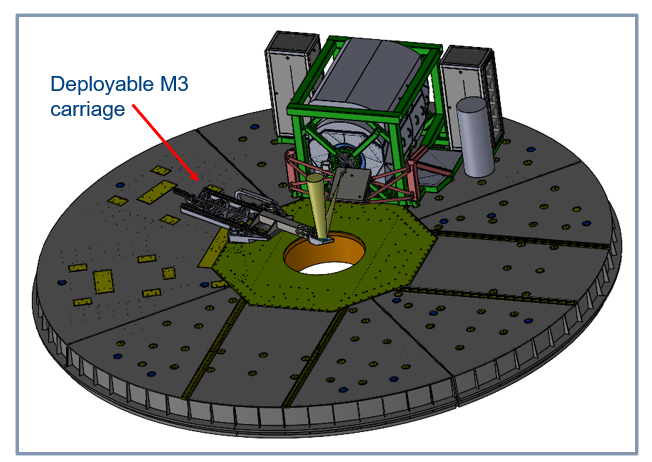}
\end{tabular}
\end{center}
\caption{Deployable M3}
\label{fig:M3} 
    \end{figure}

\begin{figure} [H]
\begin{center}
\begin{tabular}{c} 
\includegraphics[height=7cm]{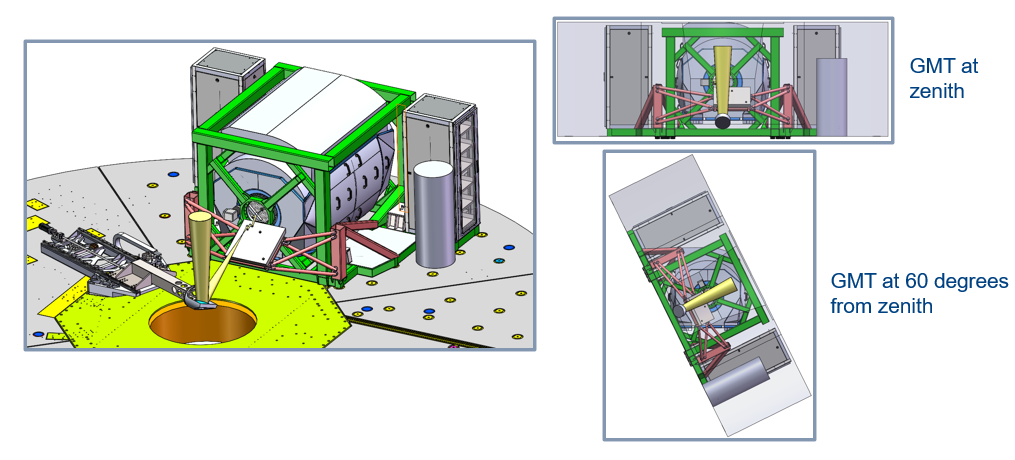}
\end{tabular}
\end{center}
\caption{GMagAO-X will be gravity invariant. The GIR will be clocked so that GMagAO-X is aligned with the elevation rotation axis when floating.}
\label{fig:gravity} 
    \end{figure}

A novel aspect of GMagAO-X is its gravity invariance (Fig.~\ref{fig:gravity}). The main bench of instrument will rotate on metal bearings in the front and back so as the telescope tilts towards a desired object for observation, the system remains upright. This is necessary due to the floating optical table used by the system.

    \begin{figure} [H]
\begin{center}
\begin{tabular}{c} 
\includegraphics[height=6cm]{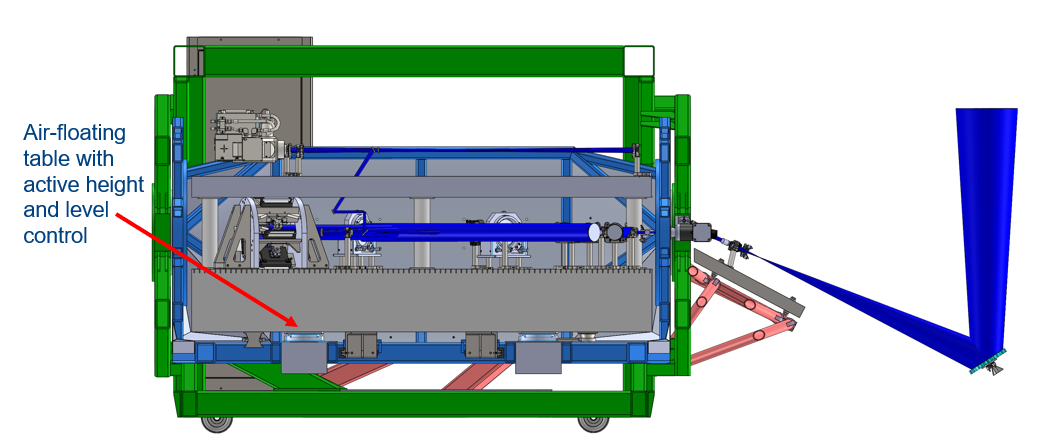}
\end{tabular}
\end{center}
\caption{Open Side View}
\label{fig:side_v3} 
    \end{figure}

\subsection{Optical Design}

The GMagAO-X optical design derives a lot of heritage from the MagAO-X optomechanical design\cite{close_optical_2018}. The optical design utilizes off-axis parabolic mirror relays for imaging between wavefront sensing and correction paths. Light from the telescope is captured by an approximately 7" x 8" elliptical M3 in a tip/tilt mount on a deployed carriage (Fig.~\ref{fig:M3}). Light is sent from the M3 up a fore-optics bench which includes a fast-steering mirror for global tip-tilt control. The light is then sent into the main gravity-invariant instrument within the rotating bearings.

    \begin{figure} [H]
\begin{center}
\begin{tabular}{c} 
\includegraphics[height=8cm]{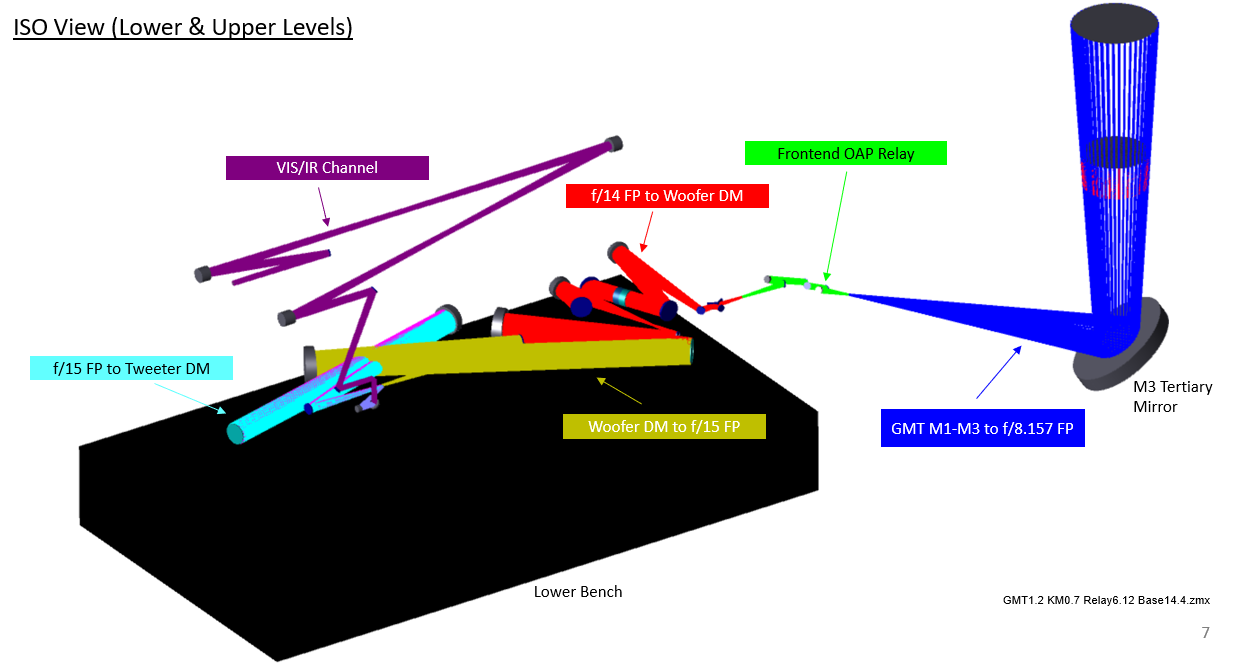}
\end{tabular}
\end{center}
\caption{Isometric View of Optical Layout}
\label{fig:iso_optics} 
    \end{figure}

    \begin{figure} [H]
\begin{center}
\begin{tabular}{c} 
\includegraphics[height=8cm]{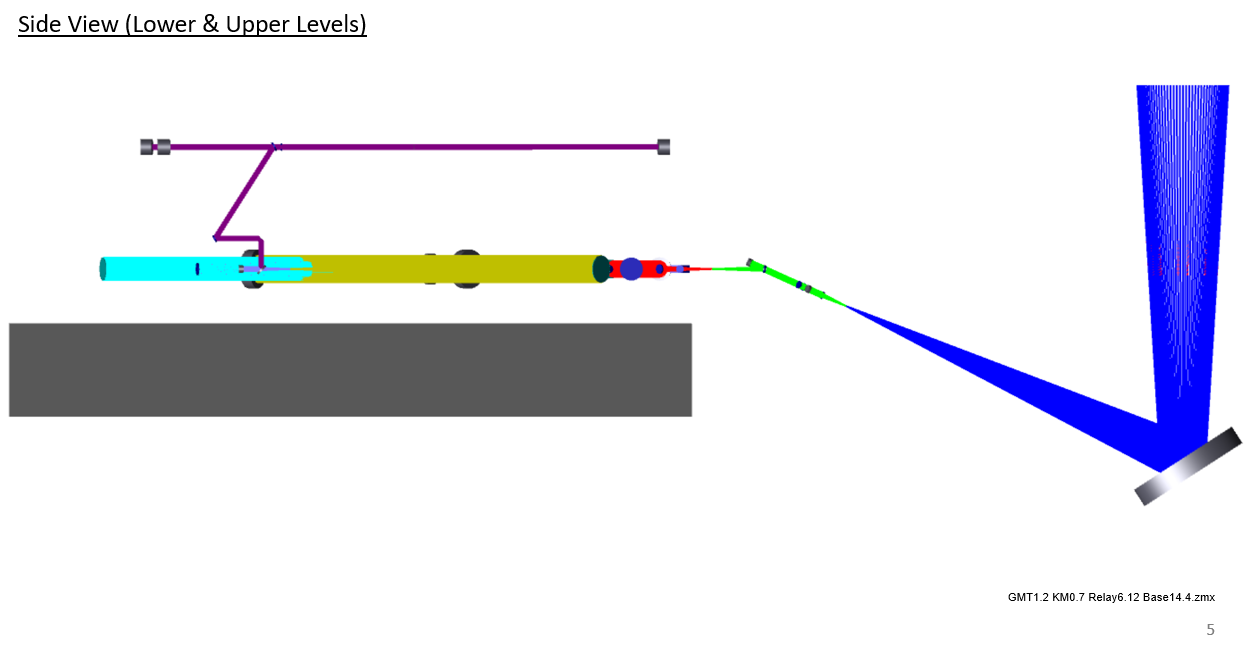}
\end{tabular}
\end{center}
\caption{Side View of Optical Layout}
\label{fig:side_optics} 
    \end{figure}

    \begin{figure} [H]
\begin{center}
\begin{tabular}{c} 
\includegraphics[height=8cm]{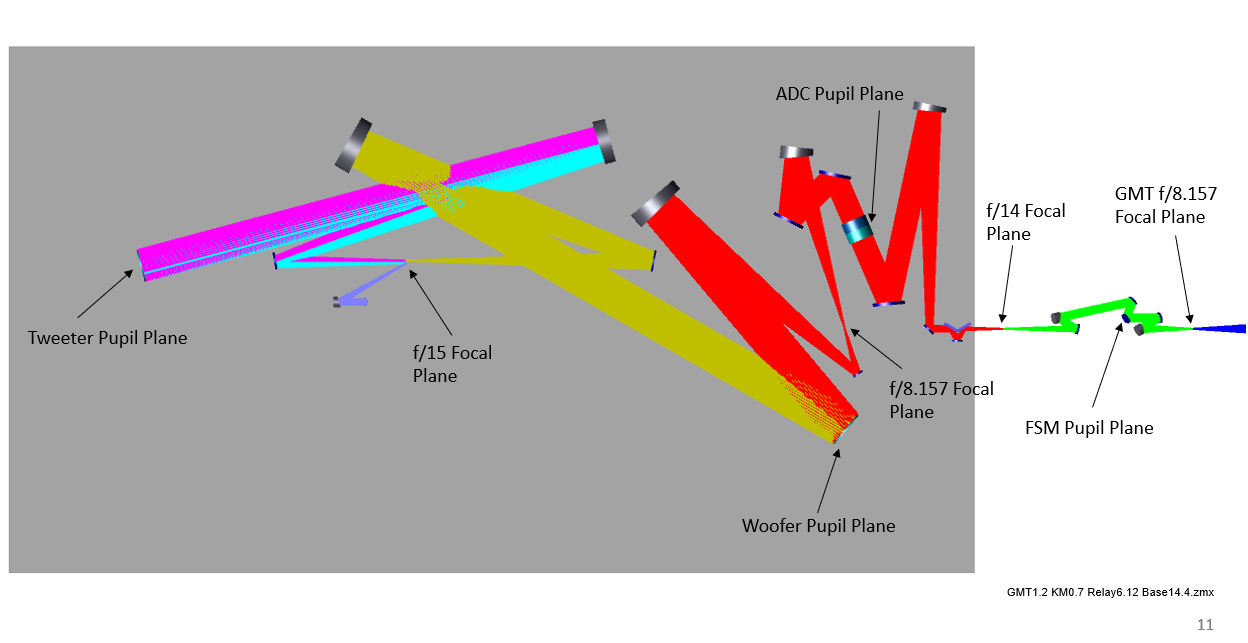}
\end{tabular}
\end{center}
\caption{Plan View of Lower Bench Optical Layout}
\label{fig:plan_lower} 
    \end{figure}


The fast-steering mirror on the fore-optics bench will allow for global tip/tilt control before the light enters the main bench.

    \begin{figure} [H]
\begin{center}
\begin{tabular}{c} 
\includegraphics[height=8cm]{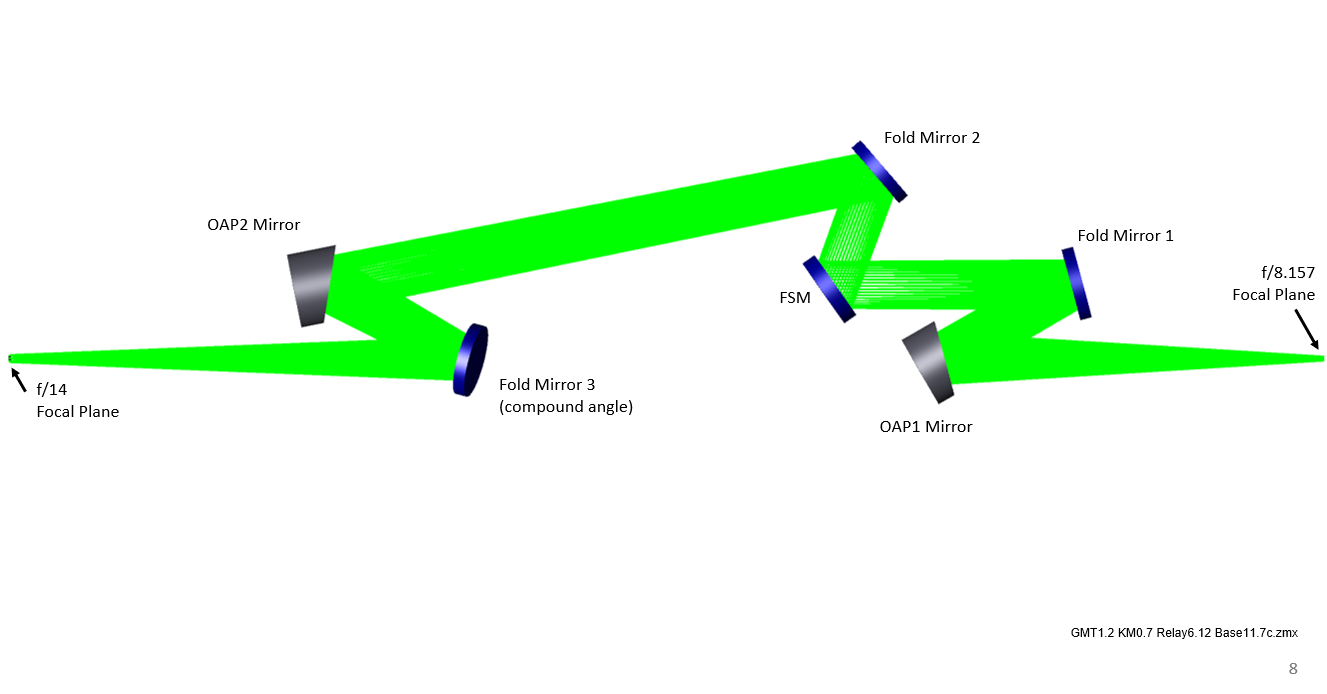}
\end{tabular}
\end{center}
\caption{Plan View of Fore-optics Optical Layout}
\label{fig:fore_top} 
    \end{figure}

\clearpage
\subsection{Optomechanical Design}
Progress has been made in selecting commercial-off-the-shelf mounts for several of the optics. Custom solutions will be designed for the remaining optics.

    \begin{figure} [H]
\begin{center}
\begin{tabular}{c} 
\includegraphics[height=8cm]{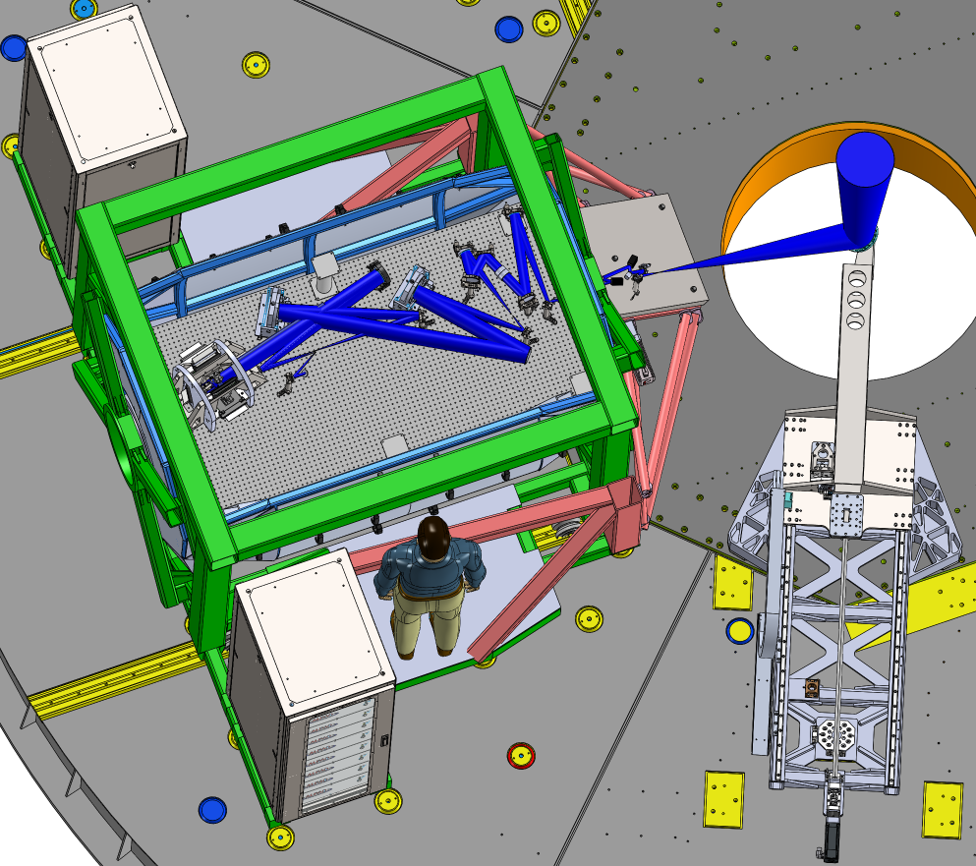}
\end{tabular}
\end{center}
\caption{Lower Bench Optics - Illustrating the adaptive optics common path going from the woofer to the parallel DM tweeter into the wavefront sensing area (wavefront sensor (PyWFS + HDFS) not shown but will occupy area on lower left of the bench).}
\label{fig:lower} 
    \end{figure}

    \begin{figure} [H]
\begin{center}
\begin{tabular}{c} 
\includegraphics[height=8cm]{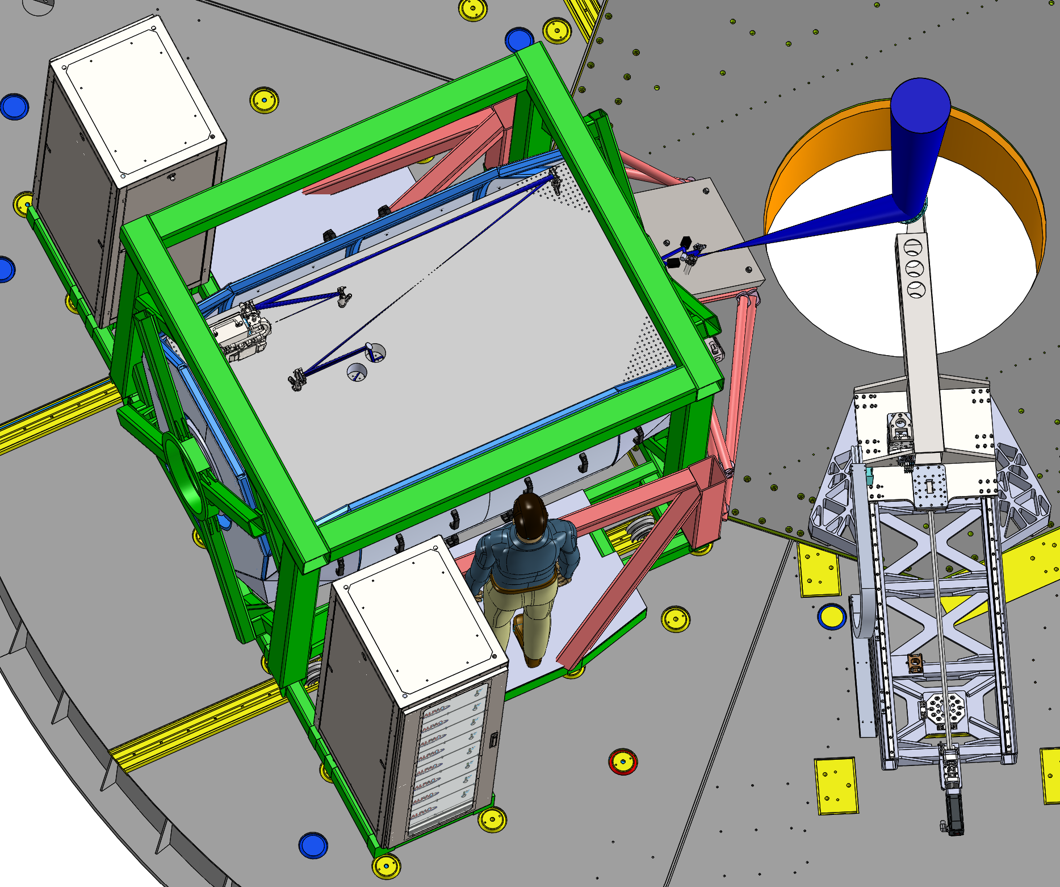}
\end{tabular}
\end{center}
\caption{Upper Bench Optics - The free space on this bench will be occupied by the post-coronagraphic science instruments, such as VIS/IR cameras and IFU spectrographs.}
\label{fig:upper} 
    \end{figure}

    \begin{figure} [H]
\begin{center}
\begin{tabular}{c} 
\includegraphics[height=9cm]{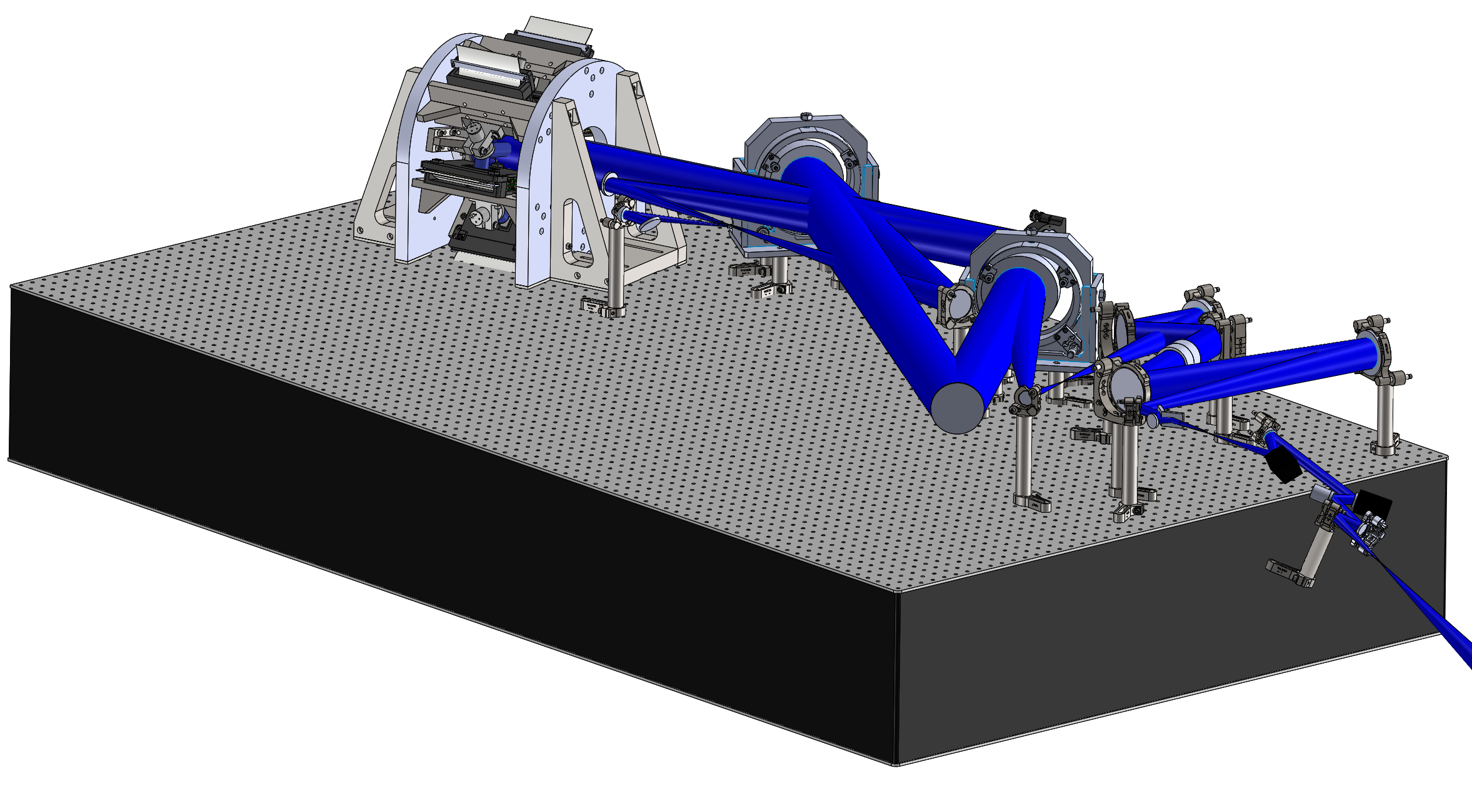}
\end{tabular}
\end{center}
\caption{Isometric View of Mounted Lower Bench Optics}
\label{fig:iso_lower_bench} 
    \end{figure}

    \begin{figure} [H]
\begin{center}
\begin{tabular}{c} 
\includegraphics[height=9cm]{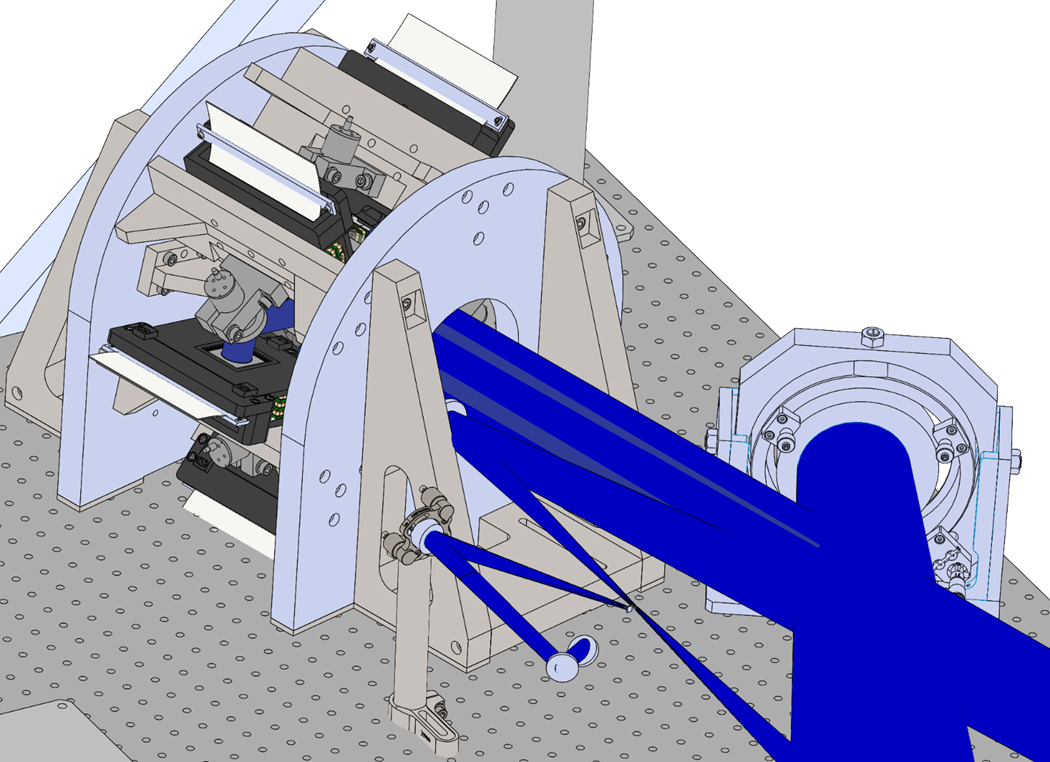}
\end{tabular}
\end{center}
\caption{Close-up of parallel DM (tweeter) mounted on table}
\label{fig:parallel_dm_again} 
    \end{figure}


\section{Testbed Development}

The U.S. Decadal Survey on Astronomy and Astrophysics 2020 stated that co-phasing all seven segments of the GMT to achieve a high-Strehl ($>$70$\%$) point spread function (PSF) would be the most significant technical challenge for the GMT project. The University of Arizona Center for Astronomical Adaptive Optics (CAAO) and the Extreme Wavefront Control Lab (XWCL) has developed the High Contrast Adaptive-optics Testbed (HCAT), to simulate co-phasing the seven GMT segments\cite{hedglen_development_2023, hedglen_first_2022, hedglen_lab_2022}.

    \begin{figure} [H]
\begin{center}
\begin{tabular}{c} 
\includegraphics[height=8cm]{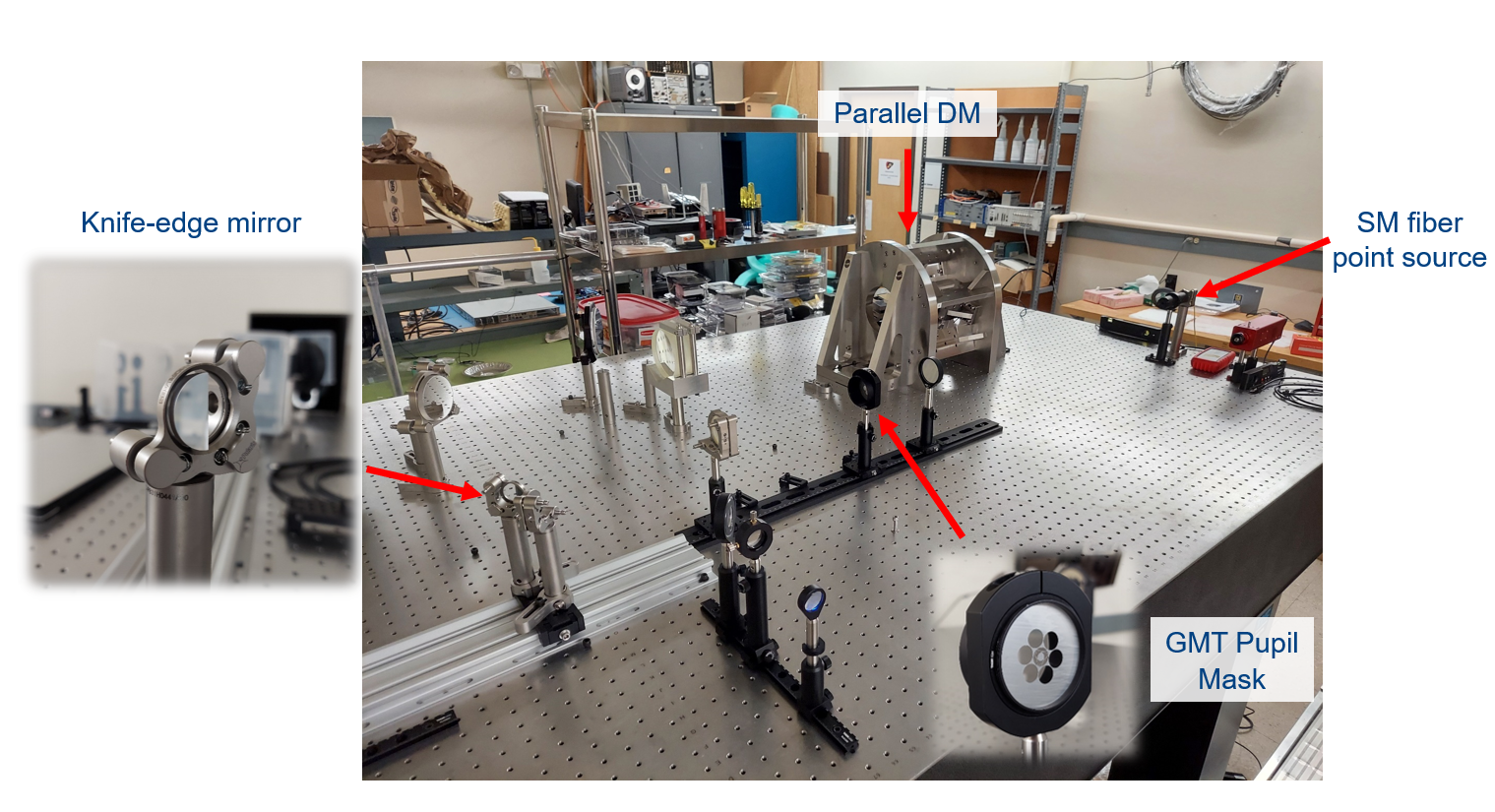}
\end{tabular}
\end{center}
\caption{HCAT built at the Steward Observatory at the University of Arizona}
\label{fig:HCAT} 
    \end{figure}

HCAT will prototype the novel parallel DM concept. In place of real DMs, the parallel DM prototype uses ``mock DMs" or simply flat mirrors mounted onto invar plates in the shape of BMC 3k DMs. This system allows HCAT to test splitting up the GMT pupil into its seven individual segments via the hexpyramid, correct piston errors via the piezoelectric PI S-325 actuators, then coherently recombine the beams. The goal is to control piston phasing errors to the tens of nanometers level.

    \begin{figure} [H]
\begin{center}
\begin{tabular}{c} 
\includegraphics[height=6cm]{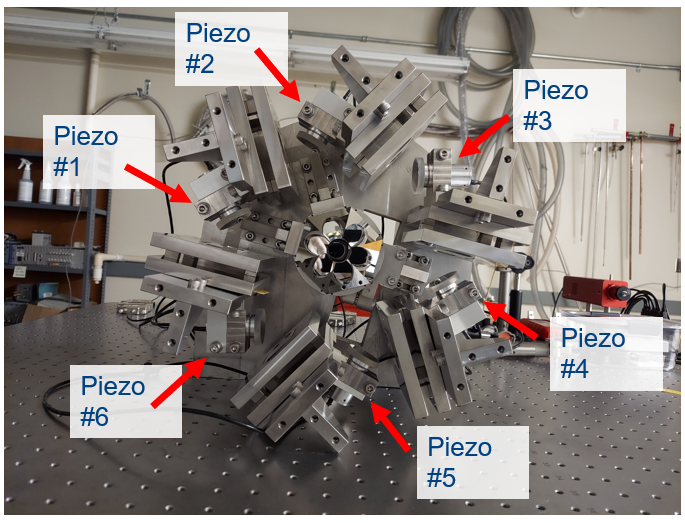}
\end{tabular}
\end{center}
\caption{Open-face manufactured parallel DM mounting structure}
\label{fig:waterwheel} 
    \end{figure}

Phasing errors are sensed by a holographic dispersed fringe sensor (HDFS) which utilizes dispersed fringes to narrow in on differential phase errors between the segments\cite{haffert_phasing_2022}.

    \begin{figure} [H]
\begin{center}
\begin{tabular}{c} 
\includegraphics[height=6cm]{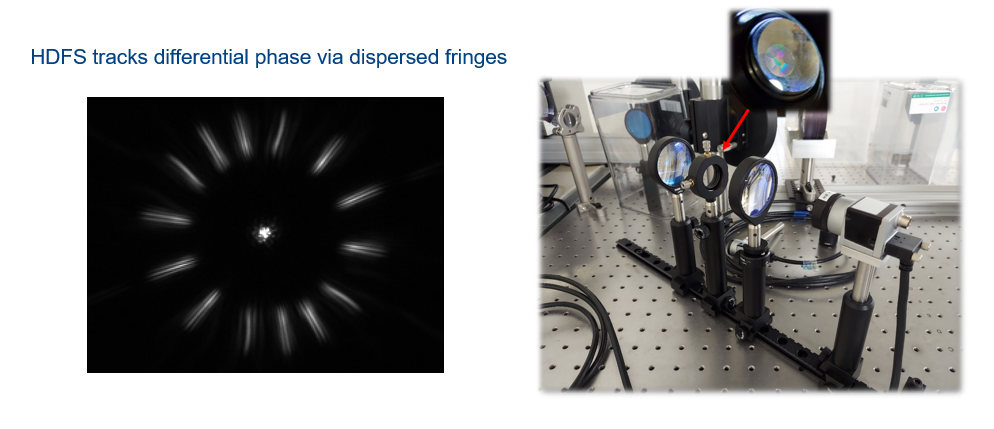}
\end{tabular}
\end{center}
\caption{(left) Pattern created by HDFS on sensor (right) HDFS on HCAT table}
\label{fig:hdfs} 
    \end{figure}

HCAT can be a standalone testbed or feed MagAO-X with an f/11 beam. The GMT pupil is relayed onto the MagAO-X 97 element woofer DM and 2040 actuator tweeter DM. This combined system allows for tests of phasing and wavefront correction on the GMT pupil, both of which are critical for the functionality of GMagAO-X.

    \begin{figure} [H]
\begin{center}
\begin{tabular}{c} 
\includegraphics[height=8cm]{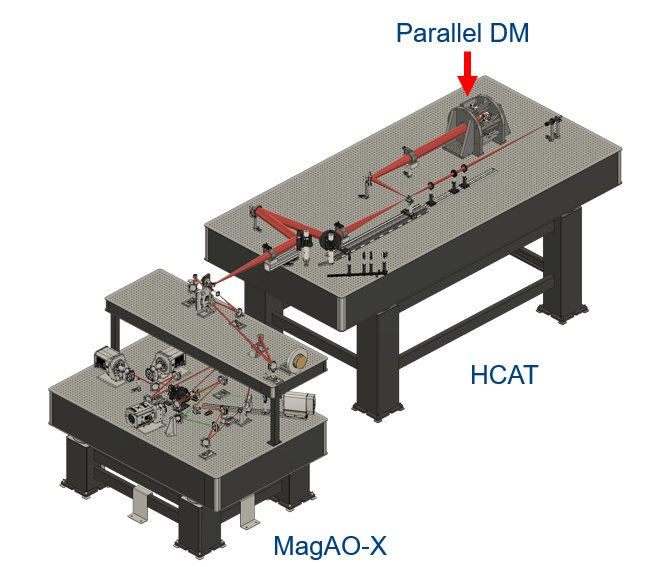}
\end{tabular}
\end{center}
\caption{CAD rendering of as-built optical feed from HCAT to MagAO-X\cite{kautz_novel_2022}}
\label{fig:feed} 
    \end{figure}

HCAT can operate in two modes: bypass mode and parallel DM mode\cite{hedglen_development_2023}. In bypass mode, two fold mirrors are placed ahead of the parallel DM to ``bypass" it and relay a perfectly phased GMT pupil (ie no splitting and recombining) to the rest of the testbed. Bypass mode is used for alignment to MagAO-X and will be utilized in training the MagAO-X woofer and tweeter to create reference PSFs. In parallel DM mode, the fold mirrors are removed and HCAT utilizes the parallel DM for phasing experiments.

    \begin{figure} [H]
\begin{center}
\begin{tabular}{c} 
\includegraphics[height=6cm]{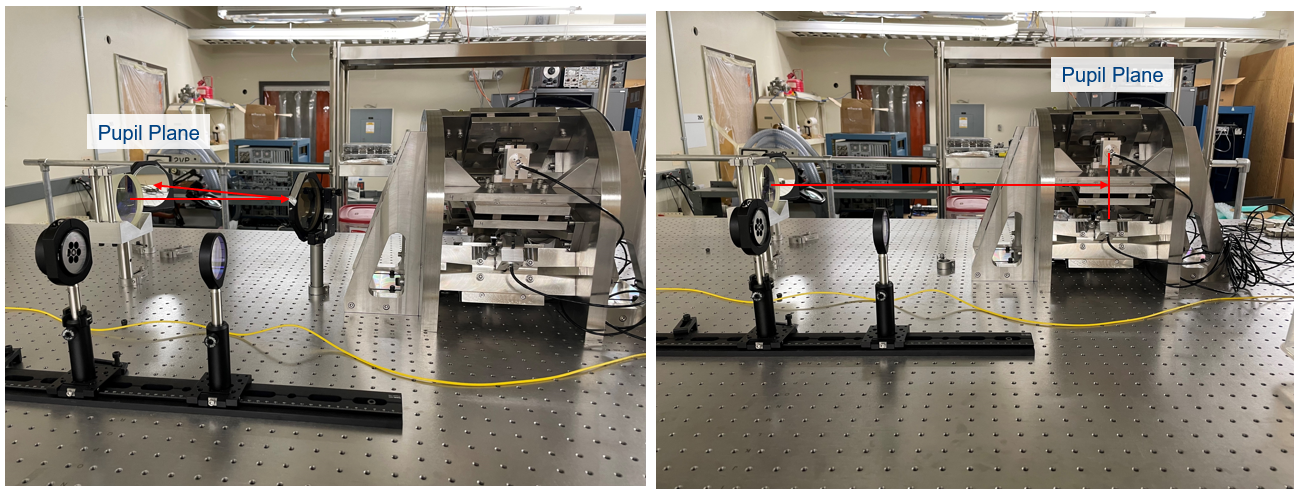}
\end{tabular}
\end{center}
\caption{(left) Bypass Mode (right) Parallel DM Mode}
\label{fig:bypass} 
    \end{figure}

\begin{figure} [H]
\begin{center}
\begin{tabular}{c} 
\includegraphics[height=7cm]{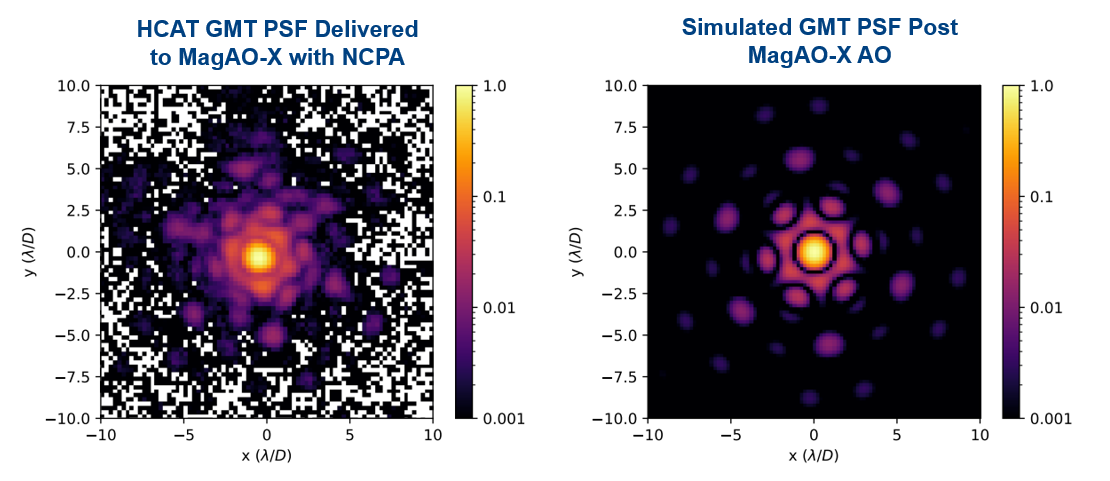}
\end{tabular}
\end{center}
\caption{(left) GMT PSF delivered by HCAT to MagAO-X with NCPA (right) Simulated GMT PSF with NCPA correction by MagAO-X}
\label{fig:HCAT_psf} 
    \end{figure}

On the left of Figure~\ref{fig:HCAT_psf} is an example of the GMT PSF delivered from HCAT in bypass mode to the MagAO-X science cameras with non-common path aberrations (NCPA). The right figure shows what the GMT PSF should look like after NCPA correction by MagAO-X.

\section{CONCLUSION}
The extreme adaptive optics system being developed for the Giant Magellan Telescope, GMagAO-X, completed a Conceptual Design Review in September 2021 and is moving towards a Preliminary Design Review in spring 2024. Near term priorities for continued development include demonstrating predictive control on-sky with MagAO-X, perfecting dark hole maintenance on-sky with MagAO-X, developing WFS telemetry-based post-processing to eliminate temporal correlations in residual atmospheric speckles, and continued AO and phasing demonstrations with HCAT. GMagAO-X will be able to characterize up to over 300 RV-known temperate and mature exoplanets in reflected light. Approximately ten of these planets are terrestrial and in the liquid water habitable zone and could possibly contain life\cite{males_conceptual_2022}. The combined power of the Giant Magellan Telescope and an extreme AO system such as this will usher in a new era of direct imaging science.

\acknowledgments 
We are very grateful for support from the NSF MRI Award \#1625441 (for MagAO-X) and funds for the GMagAO-X CoDR from the University of Arizona Space Institute. The HCAT testbed program is supported by an NSF/AURA/GMTO risk-reduction program contract to the University of Arizona (GMT-CON-04535, Task Order No. D3 High Contrast Testbed (HCAT), PI Laird Close). The authors acknowledge support from the NSF Cooperative Support award 2013059 under the AURA sub-award NE0651C. Support for this work was also provided by NASA through the NASA Hubble Fellowship grant \#HST-HF2-51436.001-A awarded by the Space Telescope Science Institute, which is operated by the Association of Universities for Research in Astronomy Inc. (AURA), under NASA contract NAS5-26555. Maggie Kautz received an NSF Graduate Research Fellowship in 2019. Alex Hedglen received a University of Arizona Graduate and Professional Student Council Research and Project Grant in February 2020. Alex Hedglen and Laird Close were also partially supported by NASA eXoplanet Research Program (XRP) grants 566 80NSSC18K0441 and 80NSSC21K0397 and the Arizona TRIF/University of Arizona “student link” program. This material is based upon work supported in part by the National Science Foundation as a subaward through Cooperative Agreement AST-1546092 and Cooperative Support Agreement AST-2013059 managed by AURA.


\printbibliography 
\end{document}